\documentclass[conference]{IEEEtran}
\IEEEoverridecommandlockouts
\usepackage{amsmath,amssymb,amsfonts}
\usepackage{graphicx}
\usepackage{textcomp}
\usepackage{xcolor}
\def\BibTeX{{\rm B\kern-.05em{\sc i\kern-.025em b}\kern-.08em
    T\kern-.1667em\lower.7ex\hbox{E}\kern-.125emX}}

\usepackage{makecell}
\usepackage{indentfirst}
\usepackage{shorttoc}

\usepackage{subfigure}
\usepackage{booktabs}
\usepackage[justification=centering]{caption}
\usepackage{algorithm}

\usepackage{multirow} 
\usepackage{algpseudocode}

\usepackage{mathrsfs}

\begin{document}

\title{TEST: an End-to-End Network Traffic Examination and Identification Framework Based on Spatio-Temporal Features Extraction
\\}

\author{\IEEEauthorblockN{Yi Zeng\IEEEauthorrefmark{1}\IEEEauthorrefmark{2}, Zihao Qi\IEEEauthorrefmark{2}, Wencheng Chen\IEEEauthorrefmark{2}$^{1}$, Yanzhe Huang\IEEEauthorrefmark{2}, Xingxin Zheng \IEEEauthorrefmark{2}, Han Qiu \IEEEauthorrefmark{3}}

\IEEEauthorblockA{\IEEEauthorrefmark{1}University of California, San Diego, CA, USA\\
 y4zeng@eng.ucsd.edu}

\IEEEauthorblockA{\IEEEauthorrefmark{2}Beijing University of Posts and Telecommunications, Beijing, China\\ 
\{qizihao,wenchengchen,yanzhe567,barathrum\}@bupt.edu.cn}

\IEEEauthorblockA{\IEEEauthorrefmark{3}Telecom Paris, Paris, France\\ 
han.qiu@telecom-paris.fr}
}

\maketitle

\begin{abstract}

With more encrypted network traffic gets involved in the Internet, how to effectively identify network traffic has become a top priority in the field. 
Accurate identification of the network traffic is the footstone of basic network services, say QoE, bandwidth allocation, and IDS. 
Previous identification methods either cannot deal with encrypted traffics or require experts to select tons of features to attain a relatively decent accuracy.
In this paper, we present a Deep Learning based end-to-end network traffic identification framework, termed $TEST$, to avoid the aforementioned problems.
CNN and LSTM are combined and implemented to help the machine automatically extract features from both special and time-related features of the raw traffic. 
The presented framework has two layers of structure, which made it possible to attain a remarkable accuracy on both encrypted traffic classification and intrusion detection tasks.
The experimental results demonstrate that our model can outperform previous methods with a state-of-the-art accuracy of 99.98\%.

\end{abstract}


\footnotetext[1]{Wencheng Chen is the corresponding author.}

\section{Introduction}
Accurate traffic classification is the footstone of basic network services, such as Quality-of-Service (QoS), Network Intrusion Detection (NID), etc \cite{zhang2015robust}. 
Attaining a high accuracy on identifying network traffic is not an easy task, however, there are obstacles lay in the way. 
For starters, security protocols like HTTPS are wieldy implemented to encrypt the network traffic \cite{qiu2019user}.
Such encryption of the network traffic brought more obfuscation to the classification and the NID task, where malicious traffic is normally encrypted \cite{bayer2009view}.
Besides, some privacy-related information, such as flow volume and flow duration won't be as easy to attain as before.
Finally, the explosive increase in the magnitude of various kinds of network traffic is challenging previous solutions with higher requirements and more demands \cite{sharma2018survey}.
Thus, a solution that can effectively identify nowadays network traffic is needed. 

With those obstacles in nowadays network environment, previous methods such as the Port Number Based method and the Data Packet Inspection (DPI) \cite{kumar2006algorithms} are outdated.
On the opposite, the methods based on Machine Learning (ML) are competent to face those obstacles. 
However, traditional ML-based methods like KNN \cite{bar2010realtime} or SVM \cite{atli2018anomaly}, their efficiency extremely rely on manually selected features, where some features are privacy-related now \cite{li2018deepnfv}.
Moreover, these methods require high storage and computational resource, which lowers the capacity of identification and adaptation \cite{zeng2018senior2local}. 
Especially for resource-constrained nodes, say smart vehicles, gateway, or cell phones, adopting a traditional ML classifier would cost way too much \cite{sakiz2017survey}. 
The other issue with traditional ML models is that most of them cannot conduct classification on the fly \cite{ye2017machine}.
Hence, a novel kind of traffic identification method that obviates those earlier mentioned obstacles is needed.

Deep Learning (DL) has also been implemented to identify network traffic.  
Taken that DL works in an end-to-end fashion, which can automatically select underlying features from raw input during training, lots of labor work can be saved \cite{wang2017end, wang2018datanet}. 
Whatsmore, DL-Based methods do not require much of computation and storage resources when conducting detection tasks.
Normally, two or three layers of convolutional neuron network can attain an accuracy exceeds the C 4.5 decision tree on identifying encrypted traffic \cite{zeng2019deep}.
Finally, DL-Based methods have a more powerful learning capability, which makes DL-Based methods easier to master complex structures and burdensome network tasks \cite{han2019efficient}. 
Thus, DL-Based solutions obviate problems involved with privacy whereas conducting high-accuracy identification tasks on the fly.

\begin{figure*}[!htb]
 \centering
 \centerline{\includegraphics[width=0.9\textwidth]{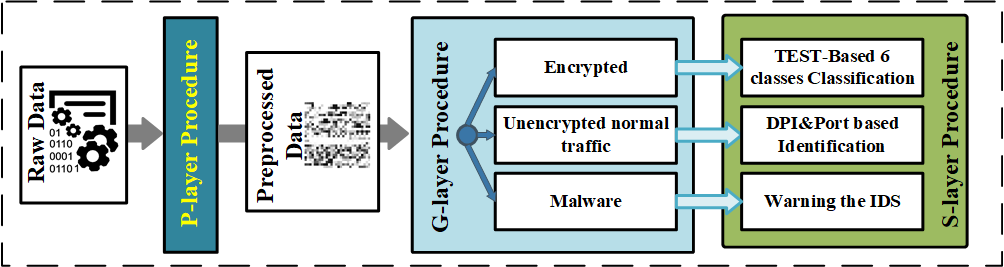}}
 \caption{Overview of the $TEST$ Framework}\label{fig.frame}
\end{figure*}

However, most presented DL-Based models adopted Convolutional Neural Network (CNN) \cite{krizhevsky2012imagenet} or Long Short-Term Memory (LSTM) \cite{hochreiter1997long} separately.
It has long been studied that CNN and LSTM can extract features from spatial and temporal domain respectively.
Thus, previously, the spatial and the temporal features of the raw traffic cannot be acquired at the same time, which limits the capability of previous methods.
For another issue, most of the previous work lose sight of the importance of using a balanced dataset during the training process \cite{wang2017hast}.
Adopting a balanced dataset can help the machine attain a higher and robust accuracy for each class in a more rapid way \cite{khan2017cost}.
Thereafter, a novel traffic classification framework based on spatial and temporal features extraction using well designed balanced datasets to train is needed.

In this paper, we present a DL-based end-to-end network traffic identification framework, termed $TEST$.
$TEST$ is the short of Traffic Examination based on Spatial and Temporal features extraction.
Thanks to the structure that adopted CNN and LSTM at the same time, $TEST$ network can use more underlying features from the raw input, hence higher accuracy is expected. 
Beside, $TEST$ framework works in a three-layered fashion, which guarantees each classifier in the framework can be trained using a separated balanced dataset.

The paper's contribution can be summarized as follows: 1)A light-weight traffic classification method named $TEST$ which utilizes both spatial and temporal features of the raw input is presented. 2)$TEST$ applies a three-layer hierarchical framework to attain a better accuracy of classification. 3)As an end-to-end framework, $TEST$ can learn from raw traffic without manual intervention and private information.
 
The rest part of this paper is organized as follows: The structure of $TEST$ and the methodology is illustrated in section II. The evaluation of this framework and the environment settings are explained in section III. The conclusion of this paper is given in section IV.

\section{THE $TEST$ FRAMEWORK}
In this section, we will explain the details of the $TEST$ framework. The $TEST$ framework can be divided into three individual functional layers as shown in Fig. \ref{fig.frame}, namely Preprocess layer (P-layer), General classification layer (G-layer), and Specialized action layer (S-layer) respectively.

The notations and parameters used in this paper are listed in TABLE \ref{tab: notionTEST}.

\begin{table}[!htb]
\setlength{\belowcaptionskip}{0.2cm}
\centering
\caption{Notations \& Parameters Used in the $TEST$}
\newcommand{\tabincell}[2]{\begin{tabular}{@{}#1@{}}#2\end{tabular}}
\begin{tabular}{c|c} 
\Xhline{1pt}
Notation & Description \\
\Xhline{1pt}
\tabincell{c} {$T$} & The total number of time units \\
\hline 
\tabincell{c} {$J$} & The total number of generated traffic-graphs\\
\hline
\tabincell{c} {$R{T^t}$} & The raw traffic of time unit $t$, $t \in (1,T)$ \\
\hline 
\tabincell{c} {$PT_i^t$} & The ${i^{th}}$ packet that captured from $R{T^t}$\\
\hline 
\tabincell{c} {$G(j)$} & The ${j^{th}}$ traffic-graph generated, $j \in (1,J)$\\
\hline 
\tabincell{c} {$Epoch$} & The number of epochs to train\\ \hline 
\tabincell{c} {$Batchsize$} & The batch size\\ 
\hline
\tabincell{c} {$\hat y$} & The output result from the classifier\\ 
\hline 
\tabincell{c} {$LearnRate$} & The learning rate of the optimizer\\ 
\hline 
\tabincell{c} {$Dropout$} & The dropout probability of the dropout process\\ 
\hline 
\tabincell{c} {$Lambda$} & The lambda parameter of the L1 regularization\\ 
\hline 
\tabincell{c} {$TrainSet$} & The training dataset used in training model\\ 
\hline 
\tabincell{c} {$S(1)$} & Warning the IDS of pertential malicious actions\\ 
\hline 
\tabincell{c} {$S(2)$} & \tabincell{c}{Further label the unencrypted normal traffic \\ based on DPI and the port number }\\ 
\hline 
\tabincell{c} {$S(3)$} & \tabincell{c}{Obtain the 6-class classifier from Algo. \ref{alg:TEST} and \\Classify the encrypted traffic into six categories}\\ 
\hline 
\Xhline{1pt}
\end{tabular}
\label{tab: notionTEST}
\end{table}

\subsection{P-layer procedure}
P-layer conducts the preprocessing procedure for the following layers, where gray-scale graphs are generated based on the raw packet input. To be specific, P-layer includes Package Generation, Traffic Purification, Traffic Refiner, Length Unification, and Numpy File Generation five different steps referring to our previous work \cite{zeng2019deep}. The difference between P-layer with our previous work's preprocessing procedure is that we store the gray-scale graphs using Numpy format, which is a more ideal input format for DL frameworks like Keras and Pytorch. 

\begin{figure}[htbp]
  \centering
  \centerline{\includegraphics[width=0.46\textwidth]{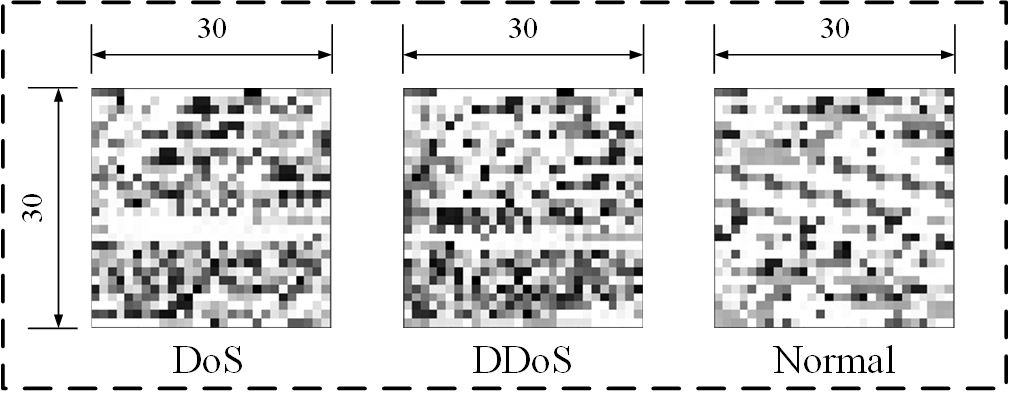}}
  \caption{Samples of Reprocessed Traffic Data}\label{fig.sample}
\end{figure}

Fig. \ref{fig.sample} is a demonstration of preprocessed traffic data using 900 bytes captured from raw packets, where bytes are mapped to pixels with values of 0 to 255. It can tell already that Malware's traffic, say DoS and DDoS shares some same features which are different from normal traffic.

\begin{figure*}[!htb]
 \centering
 \centerline{\includegraphics[width=0.98\textwidth]{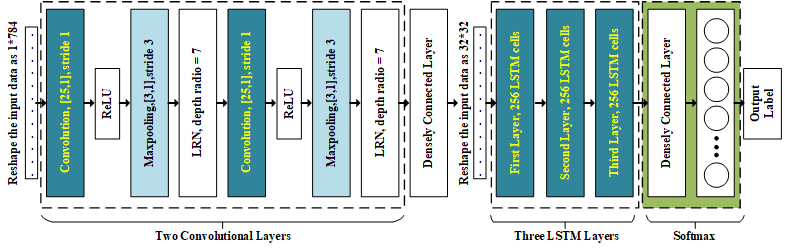}}
 \caption{Overview of the $TEST$ Model}\label{fig.structure}
\end{figure*}

\subsection{G-layer procedure}

G-layer conducts the general classification task by using the first $TEST$ model, which details are shown in Fig. \ref{fig.structure}.
Preprocessed data after going through the first $TEST$ model will be classified into three categories, Encrypted traffic, Normal traffic, and Malicious traffic respectively. The purpose of building multiple classification layers for each functional layer of $TEST$ framework is that we can use different balanced datasets to train different layers to gain better performance on each classification tasks. In other words, the proportions of Encrypted traffic, Normal traffic, and Malicious traffic in the ${G_{Train}}$  are the same.

As shown in Fig. \ref{fig.structure}, the $TEST$ model contains both CNNs and LSTMs to help the machine learn the underlying features from both spatial and temporal features.

The progress of the $TEST$ model can be summarized as follows:
For starters, we reshape the input data into $1\times784$ and feed into the first convolutional layer. 
The first convolutional layer has 32 filters with a kernel size of [1,25]. 
Then the data will go through the ReLU activation function to ensure the nonlinearity. 
After that, a Max Pooling layer is added, which kernel size is [1,3] and the stride is 3. 
The second convolutional layer has 64 filters and the rest of the settings are the same as the first one. 
After the data processed by two convolutional layers, the input data will go through a densely connected layer that has 1024 hidden units with a 50\% of dropout rate. 
Then, the input data will be reshaped as $32\times32$, which is the ideal input shape of the 3-layer LSTM structure in the $TEST$ model. 
Each layer has 256 LSTM cells, where the dropout rate is being set as 50\% to attain a better generalization. 
A densely connected layer is then added right after the 3-layer LSTM to scale done the output.  
Finally, the data will go through a Softmax classifier \cite{bouchard2007efficient} to output the predicted label. Where Softmax classifier is defined as follows:
\begin{equation}\label{EQ1}
 \setlength{\abovedisplayskip}{3pt}
 \setlength{\belowdisplayskip}{3pt}
\hat y = \frac{{\exp (Ou{t^j})}}{{\sum {\exp (Ou{t^i})} }}
\end{equation}
Where ${Ou{t^j}}$ is the output of $jth$
neuron in the densely connected layer \cite{zeng2019deep}.   
Details of the training process of the $TEST$ model is summarized in Algorithm \ref{alg:TEST}.
\begin{algorithm}[!htb]
\caption{\emph{Training Procedure of the $TEST$ Model}}
\label{alg:TEST}
\small
\begin{algorithmic}
\Require ~~\\
$TrainSet$\\ $\left\{ {Epoch,Batchsize,LR,Dropout,Lambda} \right\}$
\Ensure ~~\\
Classifier
\end{algorithmic}
\begin{algorithmic}[1]
\For{each $epoch$ in $(1,Epoch)$}
    \For{each $Batchsize$ data of the training set}
        \For{each $G$ in batch}
            \State Reshape $G$ to $1 \times 784$ form
            \State Compute convolution with 32 filters
            \State Compute the result through $ReLU$
            \State Max pooling
            \State Punish the result through the LRN layer
            \State Compute convolution with 64 filters
            \State Compute the result through $ReLU$
            \State Max pooling
            \State Punish the result through the LRN layer
            \State Run through a densely connected layer
            \State Reshape output data as 32*32
            \State Run through the 3-layered LSTM
            \State Run through a densely connected layer
            \State Output the result referring Eq. \ref{EQ1}.
            \State Update the weight \& bias. 
        \EndFor
    \EndFor
\EndFor

\end{algorithmic}
\end{algorithm}

To be noticed that during the training procedure of the $TEST$ model in the G-layer, the dataset used to train is randomly selected to be balanced. In our work, we select the same amount of encrypted traffic, unencrypted traffic, and Malware's traffic to boost accuracy in the G-layer. After a general classification conducted by the G-layer, the network traffic data will be classified into three categories, namely Encrypted traffic, Normal traffic, and Malicious traffic.

\subsection{S-layer procedure}
In S-layer, each different classes of traffic would further go through a different procedure. As for Normal traffic, by using the combination of Port number and DPI, the framework can ulteriorly classify them into more precise categories. When Malware traffic is detected, a warning will be sent through our framework to the IDS connected to warn and take further actions. Finally, for Encrypted traffic, previous methods like the methods based on the Port number and the DPI cannot precisely decode the specific class. Hence, we adopt another $TEST$ model to further classify Encrypted traffic into detailed categories. We use $TEST$ in this process to classify Encrypted traffic into six categories, Chat, File, E-mail, Stream, VoIP and P2P, which are the current network traffic divisions that are commonly used. The structure of this model is similar to the first model in the G-layer. The only difference is that we use another balanced dataset, which contains six classes of encrypted traffic with the same size.

\begin{algorithm}[!htb]
\caption{\emph{$TEST$ Framework}}
\label{alg:whole}
\small
\begin{algorithmic}
\Require ~~\\
$R{T^1},R{T^2},...,R{T^t},...,R{T^T}$
\Ensure ~~\\
$S(i) \qquad  i=1,2,3$
\end{algorithmic}
\begin{algorithmic}[1]
\For{each $t$ in $(1,T)$}
\State Split the $R{T^t}$ into $PT_1^t,PT_2^t,...PT_i^t,...$
\For{each $i$}
    \State Cut the length of $PT_i^t$ to 784 bytes 
    \State data processing on the P-layer
    \State Generate traffic-graph $G$
\EndFor
\EndFor
\State Obtain the processed data traffic-graphs $G$ from P-layer
\State Randomly select a balenced dataset from $G$
\State Obtain the 3-class classifier from Algorithm 1 in G-layer
\State Obtain the label $\hat y$ from the 3-class classifier
\If{$\hat y$ is Malware} 
\State {$S(1)$} 
\ElsIf{$\hat y$ is unencypted normal benign traffic}
\State {$S(2)$}
\ElsIf{$\hat y$ is Encrypted traffic} 
\State $S(3)$
\EndIf
\end{algorithmic}
\end{algorithm}

More details of the whole 3-layer $TEST$ framework can be summarized in Algorithm \ref{alg:whole}. In the $TEST$ framework, the raw input is the only requirement, hence the whole process can be viewed as an end-to-end structure. At the end of the $TEST$ framework, 3 kinds of actions will be conducted by the $TEST$ to treat each different class of traffic differently.

\section{Experimentation}

In this section, the details of the experimental settings are explained. Including the traffic data used for each training process, the setup of the experiment, and evaluation metrics.

\subsection{Traffic Dataset}

To fully test the effectiveness of the $TEST$ on classifying encrypted traffic and identifying malware, two public datasets are used to recreate a new dataset to simulate the environment of our case. The first dataset we select is ISCX VPN-nonVPN traffic dataset \cite{draper2016characterization}, which contains 7 kinds of encrypted traffic. To be specific, the encrypted traffic in this dataset can be divided into Chat, Email, File Transfer, P2P, Streaming, VoIP, and Browsing, which is the categories normally adopted when classifying encrypted traffic \cite{velan2015survey}. 
Since the traffic of Browsing includes traffics of all other six classes, we remove it for simplicity. The other dataset we adopt is USTC-TFC 2016 dataset. This dataset contains unencrypted benign traffic and traffic captured when malicious actions conducted. We extract 26229 samples from both benign and Malware traffic from USTC-TFC dataset respectively. The regenerated dataset hence contains both 6 classes of encrypted traffic, unencrypted benign traffic, and malware traffic. We also normalized the size of each class to train models faster. We randomly select 20984 samples of encrypted traffic, normal benign, and malicious traffic respectively for the training process of the 3-class $TEST$.
As for the 6-class classification task, we only use 818 samples of each class in the training process to attain a balanced shape.
Other details of the dataset that we used in this paper are summarized in Table \ref{tab:struc}.

\begin{table}[!htb]
\setlength{\belowcaptionskip}{0.2cm}
\centering
\caption{Structure of the Balenced Dataset for 3-class}
\newcommand{\tabincell}[2]{\begin{tabular}{@{}#1@{}}#2\end{tabular}}
\begin{tabular}{ p{1.5cm}<{\centering} |p{0.925cm}<{\centering} |p{1.869cm}<{\centering}} 
\Xhline{1pt}
Class Name & Quantity & Percentage (\%) \\
\hline
\tabincell{c} Chat & 5840 & 7.42\\
\hline 
\tabincell{c} Email & 5852 & 7.44\\
\hline 
\tabincell{c} File & 5839 & 7.42\\
\hline 
\tabincell{c} P2P & 1022 & 1.30\\
\hline 
\tabincell{c} Streaming & 1829 & 2.32\\
\hline 
\tabincell{c} VoIP & 5847 & 7.43\\
\hline 
\tabincell{c} Benign & 26229 & 33.33\\
\hline 
\tabincell{c} Malware & 26229 & 33.33\\ 
\Xhline{1pt}
\tabincell{c} TOTAL & 78687 & 100\\
\Xhline{1pt}
\end{tabular}
\label{tab:struc}
\end{table}

\subsection{Experimental Setup}
Tensorflow is used as the experiment ML software framework. We run the evaluation on Ubuntu 18.04 64 bit OS. The processor is an 8 cores Intel I7-7700K CPU with 32 GB of memory. Two chips of Nvidia GeForce GTX 2080 Ti are used as the GPU accelerators.

The hyperparameters in the training process of the $TEST$ model is set as follows: We train each $TEST$ model using 5000 $Epoch$ with $Batchsize$ being set as 200. The $LearnRate$ of the Adam optimizer is set as 0.0006. The $Dropout$ is being set as 50\%. Finally, the $Lambda$ of the L1 regularization in the convolutional layers and LSTM layers are being set as 0.0005 and 0.00009 respectively.

To be noticed that the comparison group, namely the LeNet and the 3-layer LSTM's hyperparameters are not the same as the $TEST$ model. Due to the difference in the structure of each network, we fine-tune LeNet and the 3-layer LSTM using different parameters that fit with them best.

\subsection{Evaluation Metrics}
After the training process of two $TEST$  models in two layers, we use the testing dataset to evaluate the effectiveness of the $TEST$ framework.
We first compare the accuracy of our presented framework with LeNet and 3-layer LSTM, where the accuracy is defined as follows:
\begin{equation}\label{EQ2}
 \setlength{\abovedisplayskip}{3pt}
 \setlength{\belowdisplayskip}{3pt}
Accuracy = \frac{TP+TN}{TP+FP+FN+TN}
\end{equation}
where TP is True Positive, namely the number of correctly classified cases as a specific class; FP is False Positive,namely the number of misclassified cases that classified as
that class; FN, False Negative, which is the number of cases that are supposed to be classified as that class, yet misclassified as other classes; TN, True Negative, which is the number of cases that correctly classified as not that specific class.

\begin{figure*}[!htb]
 \centering
 \centerline{\includegraphics[width=0.98\textwidth]{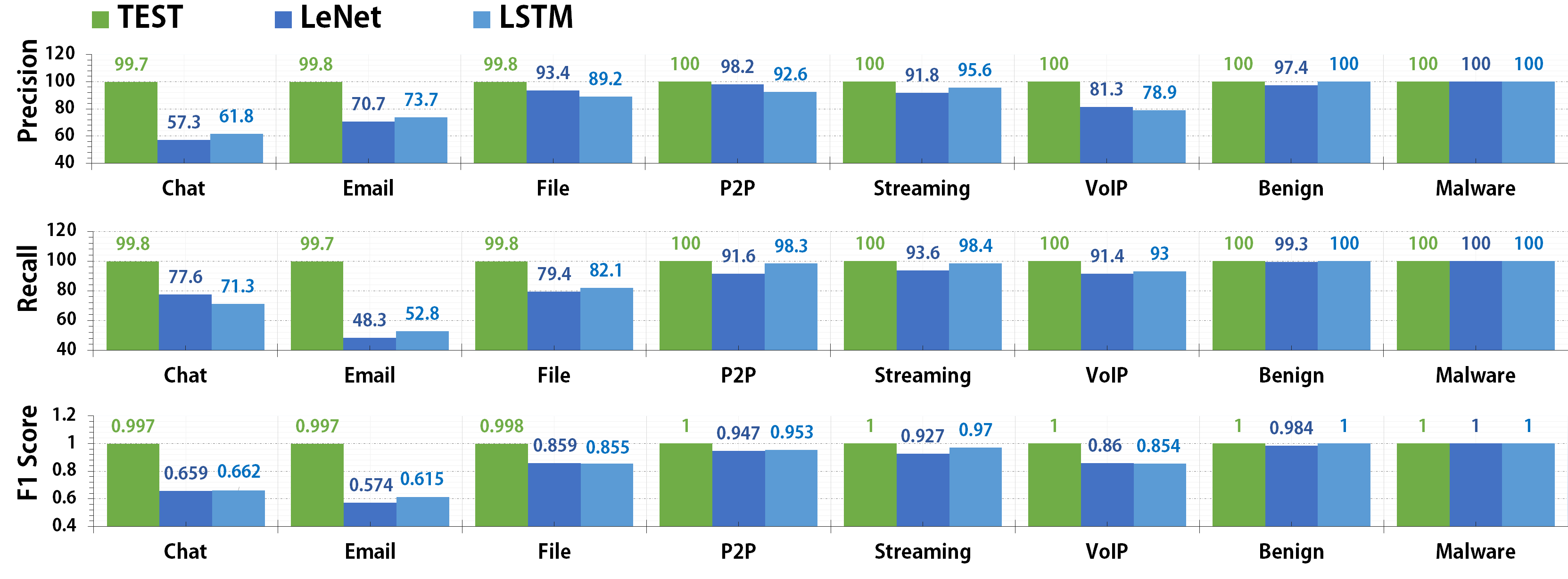}}
 \caption{Comparison of each Method}\label{fig.Compar}
\end{figure*}

We then evaluate and compare the performance of the $TEST$ with the other two networks using three metrics. Namely, Precision, Recall, and F1 score, which are defined as follows:
\begin{equation}\label{EQ2}
 \setlength{\abovedisplayskip}{3pt}
 \setlength{\belowdisplayskip}{3pt}
Precision = \frac{TP}{TP+FP}
\end{equation}
\begin{equation}\label{EQ2}
 \setlength{\abovedisplayskip}{3pt}
 \setlength{\belowdisplayskip}{3pt}
Recall = \frac{TP}{TP+FN}
\end{equation}
\begin{equation}\label{EQ2}
 \setlength{\abovedisplayskip}{3pt}
 \setlength{\belowdisplayskip}{3pt}
F1\_score = \frac{2 \times Precision\times Recall}{Precision+Recall}
\end{equation}
Those three metrics are wieldy used in various classification tasks to evaluate how effective a method can be.

\section{Evaluation}

In this section, we compare the performance of $TEST$ with other DL-Based models, namely the LeNet and 3-layer LSTM. LeNet is based on 2D CNN with L2 regularization, which in our case is used as a comparison group that uses only spatial characteristics of the raw traffic. LSTM otherwise, is used as a comparison group which only deploy the temporal features from the raw input.

The accuracy of the three models on the testing dataset is shown in Table \ref{tab: acc}. The comparations of testing evaluations are shown in Figure \ref{fig.Compar}.

\begin{table}[!htb]
 \setlength{\belowcaptionskip}{0.1cm}

 \centering
 \caption{Accuracy of different models on traffic.}
 \newcommand{\tabincell}[2]{\begin{tabular}{@{}#1@{}}#2\end{tabular}}
 \begin{tabular}{c|c} 
 \Xhline{1pt}
Model & Accuracy \\
 \Xhline{1pt}
 \tabincell{c} {${TEST}$} & 99.98\% \\
 \hline 
 \tabincell{c} {${LeNet}$} & 80.27\% \\
 \hline
 \tabincell{c} {${LSTM}$} & 81.96\% \\
 \hline 
 \Xhline{1pt}
 \end{tabular}
 \label{tab: acc}
\end{table}

As we can tell from Table \ref{tab: acc}, the $TEST$ can attain the highest accuracy — 99.98\% on the network traffic classification task that attributes to its uses both CNN and RNN. As a comparison that only extracts spatial features, the LeNet attains an accuracy of 80.27\%.
Meanwhile, 3-layer LSTM, which utilizes temporal features to attain an accuracy of 81.96\%. 
Based on such a result, as for the traffic data, temporal features might contain more decisive information to help the machine to learn than spatial features.
It is worth to mention that, the TEST framework ensures every specific task uses a designed balanced dataset, hence capable to attain an accuracy almost 100\%.
The result of accuracy confirmes with our expectation on utilizing spatial and temporal features at the same time will boost the efficiency of the method.

Then, the effectiveness of the three models toward each specific class is presented in Fig. \ref{fig.Compar}. We compare the precision, recall, and F1 score of above three models toward each class to draw a detailed observation.


From Fig. \ref{fig.Compar},  we can observe that the $TEST$ attains a better result and robustness than LeNet and LSTM on the testing dataset. 
First, the precision and recall of the $TEST$ take the lead, especially in classifying Chat, Email, and File transferring traffic comparing to the other two DL-Based methods. For the rest of the classes of traffic, $TEST$ attains a high performance with an F1 score of 1. The outstanding capability on utilizing bothe spatial and temporal characters with balanced dataset deployed helps $TEST$ framework dominate in classifying and identifying the aforementioned 5 classes. As a contrast, extracting features from ether just spatial range or just temporal domain would lose the robustness toward classifying different classes.

It is worth to mention that LeNet and LSTM somewhat show high and stable results when classifying unencrypted Begin traffic and Malware's traffic, which is almost as good as the $TEST$ framework. Such a result is due to that unencrypted Begin traffic has more stabilized features which are much more easy to be learned by the machine than the encrypted traffic. Similarly, malicious traffic in our case tries to concealed as normally unencrypted traffic, hence features are easier to extract by DL-Based models than encrypted traffic.

Moreover, the average F1 Score of  $TEST$, LeNet and LSTM are 0.999, 0.851, 0.864 respectively. $TEST$'s performance is more effective and robustness compared with the LeNet and the 3-layer LSTM. That is to say, the performance of $TEST$ in classifying traffic is not only good but also stable. It demonstrates that $TEST$ can give an exact classification of traffic data and attains higher robustness. 

It can be summarized that merely extract underlying features from one dimension would highly constrain the accuracy and robustness of a DL-Based model towards traffic data. In brief, $TEST$ outperforms the other two models on traffic classification tasks.

\section{Future Work and Conclusion}

Deep Learning (DL) models have recently found vulnerable to Adversarial Examples (AEs), which become a potential threat to DL models implemented in safety-concerning places. AEs are a kind of sample that generated based on original authentic data, which the difference is imperceptible to humans, yet can cause a DL model to label the image incorrectly \cite{Nguyen_2015_CVPR}. Due to the conditional constraint, we did not discuss the robustness of our TEST framework toward various kind of crafted AEs. TEST can detect Malware's traffic and classify encrypted traffic at a state-of-the-art level. However, if the robustness toward AEs is not ideal, the framework might cause sever consequence. In future work, we will develop several network traffic AEs generating framework to test the robustness of the proposed TEST. Meanwhile, by exploring ways to enhance the robustness of our proposed frameworks is also a future direction.

In this paper, we presented a Deep Learning based end-to-end network traffic identification framework,  termed $TEST$. Compared to previous DL-Based methods, $TEST$ can extract both spatial and temporal features from the raw input. A hierarchical structure was adopted to ensure each $TEST$ model in the framework can attain a fast and robust training using balanced datasets.
Thanks to this unique structure and the help from both CNN and LSTM,  $TEST$ attained a state-of-the-art accuracy of 99.98\% on a dataset contains 8 class of traffic. From the experiment, it has been proved that $TEST$ framework is competent enough to deal with real-life traffic identification tasks.





\bibliographystyle{IEEEtran}
\bibliography{thebibfile}

\begin{thebibliography}{10}
\providecommand{\url}[1]{#1}
\csname url@samestyle\endcsname
\providecommand{\newblock}{\relax}
\providecommand{\bibinfo}[2]{#2}
\providecommand{\BIBentrySTDinterwordspacing}{\spaceskip=0pt\relax}
\providecommand{\BIBentryALTinterwordstretchfactor}{4}
\providecommand{\BIBentryALTinterwordspacing}{\spaceskip=\fontdimen2\font plus
\BIBentryALTinterwordstretchfactor\fontdimen3\font minus
  \fontdimen4\font\relax}
\providecommand{\BIBforeignlanguage}[2]{{%
\expandafter\ifx\csname l@#1\endcsname\relax
\typeout{** WARNING: IEEEtran.bst: No hyphenation pattern has been}%
\typeout{** loaded for the language `#1'. Using the pattern for}%
\typeout{** the default language instead.}%
\else
\language=\csname l@#1\endcsname
\fi
#2}}
\providecommand{\BIBdecl}{\relax}
\BIBdecl

\bibitem{zhang2015robust}
J.~Zhang, X.~Chen, Y.~Xiang, W.~Zhou, and J.~Wu, ``Robust {N}etwork {T}raffic
  {C}lassification,'' \emph{IEEE/ACM Transactions on Networking (TON)},
  vol.~23, no.~4, pp. 1257--1270, 2015.

\bibitem{qiu2019user}
H.~Qiu, H.~Noura, M.~Qiu, Z.~Ming, and G.~Memmi, ``A {U}ser-{C}entric {D}ata
  {P}rotection {M}ethod for {C}loud {S}torage {B}ased on {I}nvertible {DWT},''
  \emph{IEEE Transactions on Cloud Computing}, 2019.

\bibitem{bayer2009view}
U.~Bayer, I.~Habibi, D.~Balzarotti, E.~Kirda, and C.~Kruegel, ``A {V}iew on
  {C}urrent {M}alware {B}ehaviors.'' in \emph{LEET}, 2009.

\bibitem{sharma2018survey}
S.~Sharma and A.~Kaul, ``A survey on {I}ntrusion {D}etection {S}ystems and
  {H}oneypot based proactive security mechanisms in {VANET}s and {VANET}
  {C}loud,'' \emph{Vehicular Communications}, 2018.

\bibitem{kumar2006algorithms}
S.~Kumar, S.~Dharmapurikar, F.~Yu, P.~Crowley, and J.~Turner, ``Algorithms to
  accelerate multiple regular expressions matching for deep packet
  inspection,'' in \emph{ACM SIGCOMM Computer Communication Review}, vol.~36,
  no.~4, 2006, pp. 339--350.

\bibitem{bar2010realtime}
R.~Bar-Yanai, M.~Langberg, D.~Peleg, and L.~Roditty, ``Realtime
  {C}lassification for {E}ncrypted {T}raffic,'' in \emph{International
  Symposium on Experimental Algorithms}, 2010, pp. 373--385.

\bibitem{atli2018anomaly}
B.~G. Atli, Y.~Miche, A.~Kalliola, I.~Oliver, S.~Holtmanns, and A.~Lendasse,
  ``Anomaly-based intrusion detection using extreme learning machine and
  aggregation of network traffic statistics in probability space,''
  \emph{Cognitive Computation}, vol.~10, no.~5, pp. 848--863, Oct 2018.

\bibitem{li2018deepnfv}
L.~Li, K.~Ota, and M.~Dong, ``Deepnfv: A light-weight framework for intelligent
  edge network functions virtualization,'' \emph{IEEE Network}, no.~99, pp.
  1--6, 2018.

\bibitem{zeng2018senior2local}
Y.~Zeng, M.~Qiu, Z.~Ming, and M.~Liu, ``Senior2local: A machine learning based
  intrusion detection method for vanets,'' in \emph{International Conference on
  Smart Computing and Communication}, 2018, pp. 417--426.

\bibitem{sakiz2017survey}
F.~Sakiz and S.~Sen, ``A survey of attacks and detection mechanisms on
  intelligent transportation systems: {VANET}s and {I}o{V},'' \emph{Ad Hoc
  Networks}, vol.~61, pp. 33--50, 2017.

\bibitem{ye2017machine}
H.~Ye, L.~Liang, G.~Y. Li, J.~Kim, L.~Lu, and M.~Wu, ``Machine learning for
  vehicular networks,'' \emph{arXiv preprint arXiv:1712.07143}, 2017.

\bibitem{wang2017end}
W.~Wang, M.~Zhu, J.~Wang, X.~Zeng, and Z.~Yang, ``End-to-end encrypted traffic
  classification with one-dimensional convolution neural networks,'' in
  \emph{Intelligence and Security Informatics (ISI), 2017 IEEE International
  Conference on}, 2017, pp. 43--48.

\bibitem{wang2018datanet}
P.~Wang, F.~Ye, X.~Chen, and Y.~Qian, ``Datanet: Deep learning based encrypted
  network traffic classification in sdn home gateway,'' \emph{IEEE Access},
  vol.~6, pp. 55\,380--55\,391, 2018.

\bibitem{zeng2019deep}
Y.~Zeng, H.~Gu, W.~Wei, and Y.~Guo, ``${D}eep-{F}ull-{R}ange$: a {D}eep
  {L}earning {B}ased {N}etwork {E}ncrypted {T}raffic {C}lassification and
  {I}ntrusion {D}etection {F}ramework,'' \emph{IEEE Access}.

\bibitem{han2019efficient}
H.~Qiu, M.~Qiu, Z.~Lv, and G.~MEMMI, ``An {E}fficient {K}ey {D}istribution
  {S}ystem for {D}ata {F}usion in {V}2{X} {H}eterogeneous {N}etworks,''
  \emph{Information Fusion}, 2019.

\bibitem{krizhevsky2012imagenet}
A.~Krizhevsky, I.~Sutskever, and G.~E. Hinton, ``Imagenet classification with
  deep convolutional neural networks,'' in \emph{Advances in neural information
  processing systems}, 2012, pp. 1097--1105.

\bibitem{hochreiter1997long}
S.~Hochreiter and J.~Schmidhuber, ``Long short-term memory,'' \emph{Neural
  computation}, vol.~9, no.~8, pp. 1735--1780, 1997.

\bibitem{wang2017hast}
W.~Wang, Y.~Sheng, J.~Wang, X.~Zeng, X.~Ye, Y.~Huang, and M.~Zhu, ``Hast-ids:
  Learning hierarchical spatial-temporal features using deep neural networks to
  improve intrusion detection,'' \emph{IEEE Access}, vol.~6, pp. 1792--1806,
  2017.

\bibitem{khan2017cost}
S.~H. Khan, M.~Hayat, M.~Bennamoun, F.~A. Sohel, and R.~Togneri,
  ``Cost-sensitive learning of deep feature representations from imbalanced
  data,'' \emph{IEEE transactions on neural networks and learning systems},
  vol.~29, no.~8, pp. 3573--3587, 2017.

\bibitem{bouchard2007efficient}
G.~Bouchard, ``Efficient bounds for the softmax function and applications to
  approximate inference in hybrid models,'' in \emph{NIPS 2007 workshop for
  approximate Bayesian inference in continuous/hybrid systems}, 2007.

\bibitem{draper2016characterization}
G.~Draper-Gil, A.~H. Lashkari, M.~S.~I. Mamun, and A.~A. Ghorbani,
  ``Characterization of encrypted and vpn traffic using time-related,'' in
  \emph{Proceedings of the 2nd international conference on information systems
  security and privacy (ICISSP)}, 2016, pp. 407--414.

\bibitem{velan2015survey}
P.~Velan, M.~{\v{C}}erm{\'a}k, P.~{\v{C}}eleda, and M.~Dra{\v{s}}ar, ``A survey
  of methods for encrypted traffic classification and analysis,''
  \emph{International Journal of Network Management}, vol.~25, no.~5, pp.
  355--374, 2015.

\bibitem{Nguyen_2015_CVPR}
A.~Nguyen, J.~Yosinski, and J.~Clune, ``Deep neural networks are easily fooled:
  High confidence predictions for unrecognizable images,'' in \emph{The IEEE
  Conference on Computer Vision and Pattern Recognition (CVPR)}, June 2015.

\end{thebibliography}

\end{document}